# Evaluation of Road User Radio-Frequency Exposure Levels in an Urban Environment from Vehicular Antennas and the Infrastructure in ITS-G5 5.9 GHz Communication

**Martina Benini[1,2], Silvia Gallucci[2], Marta Bonato[2], Marta Parazzini[2], Gabriella Tognola[2]**
[1]Dipartimento di Elettronica, Informazione e Bioingegneria (DEIB), Politecnico di Milano, 20133, Milan, Italy.
[2]Istituto di Elettronica e di Ingegneria dell'Informazione e delle Telecomunicazioni (IEIIT), Consiglio Nazionale delle Ricerche (CNR), 20133, Milan, Italy.

Corresponding author: Martina Benini (e-mail: martinabenini@cnr.it)

**Abstract**
This study aims to investigate the variability of exposure levels among road users generated in a realistic urban scenario by Vehicle-to-Vehicle (V2V) and Vehicle-to-Infrastructure (V2I) communication technologies operating at 5.9 GHz. The exposure levels were evaluated in terms of whole-body Specific Absorption Rate (wbSAR) [W/kg] in three different human models, ranging from children to adults. We calculated the electromagnetic field exposure level generated by V2V and V2I using raytracing and we assessed wbSAR resulting in urban exposure scenarios with an increasing number of transmitting antennas. Whole-body SAR was generally very low, on the order of $10^{-4}$ W/kg. The maximum wbSAR, of $4.9 \cdot 10^{-4}$ W/kg, was obtained in the worst-case exposure condition comprising more than one transmitting vehicle and was found in the adult model for a distance within 10 m from the transmitting cars. We found that the height of the human model highly impacted the exposure level. Namely, the child (which is the shortest human model) was generally much less exposed than adults. All the wbSAR values found by varying the number of transmitting antennas, the distance of the road user from the antennas, and the type of human model (adult vs. child) were very well below the limits set by the ICNIRP and IEEE guidelines of 0.08 W/kg for human exposure in the 100 kHz – 300 GHz range.

## 1. Introduction

The rapid expansion of cities, particularly metropolitan areas, has resulted in an increase in traffic flux in the present era, which, according to the World Health Organization (WHO), leads to 1.19 million people deaths in Europe every year resulting from road accidents [1]. In addition, air pollution caused by car emissions has been a major environmental issue. To address these problems, numerous research endeavors are dedicated to achieving the implementation of a new paradigm called Vehicular Ad-Hoc Networks (VANETs). This paradigm plays a crucial role in the evolution of the new concept of Intelligent Transport System (ITS) [2], aimed at elevating the overall quality of road user life and traffic conditions. Among the many technologies embedded in the ITS, Vehicle-to-Everything (V2X) communication is the backbone for connected and autonomous vehicles on the road, creating a wireless network where vehicles, infrastructures, and pedestrians exchange information. These technologies are specifically referred to Vehicle-to-Vehicle (V2V), Vehicle-to-Infrastructure (V2I), Vehicle-to-Pedestrian (V2P) and Vehicle-to-Network (V2N) communication [3][4][5][6]. More precisely, V2V provides the communication between vehicles through On-Board Units (OBUs), i.e., installations within the vehicle designed for information transmission; V2I provides the exchange of information with infrastructural Roadside Units (RSUs), and V2P involves communication with electric devices owned by the road user such as smartphones or generic wearables devices. V2X technologies are mainly based on two major wireless access standards, i.e., *i)* Dedicate Short-Range Communication (DSCR) based on the well-established IEEE 802.11p that operates at 5.9 GHz [7] and *ii)* Cellular-V2X (C-V2X) [8][9][10][11][12] which is a more recent technology that expands the functionality of V2X via the use of 5G technologies with the aim to improve the performance of the communication systems minimizing at the same time the radiation spread in the environment. However, with the advent of all the V2X technologies people inside and outside the vehicles will be exposed to many RF Electromagnetic fields (RF-EMF). Besides the many articles investigating the technical aspect of the V2X wireless wave propagation, such as improvement and reliability in the exchange of the signal information [13][14][15], few articles in the literature investigated the RF-EMF exposure levels on human body generated by these V2X technologies. In particular, the authors in [16][17][18] investigated with a deterministic approach the RF-EMF exposure on a road user inside [16] and outside [17][18] a car equipped with V2V antennas operating at 5.9 GHz. In [19][20] the authors investigated the exposure levels on a road user generated by V2V technology at 3.5 GHz (considering the C-V2X protocol) with both deterministic [19] and stochastic methods [20]. Finally, the authors in [21] investigated the dose absorbed generated by the V2V technology at 5.9 GHz on a road user considering additional factors to mimic an urban layout composed of buildings and roads. In all these studies there was evidence that the dose absorbed by the human body was always below the basic restriction of the ICNIRP [22] and IEEE [23] guidelines of 0.08 W/kg in the overall body, and 2 W/kg in 10 g of tissues in the head and torso region, and 4 W/kg in 10 g tissues of the limb region.

In contrast to the previous articles that examined the exposure levels of road users in free space [16][17][18][19][20] and, in the best case, using an analytical approach that accounts for the influence of buildings and roads on the propagation of the RF field generated by vehicular antennas [21], this current study wants to consider a more realistic urban scenario that includes 3D models of all the most characteristic features seen in a real urban environment, i.e., buildings, roads and vegetation like trees and grass. Furthermore, while most of the attention in the previous studies was placed on V2V technology [16][17][18][19][20][21], nothing is known about the exposure levels emitted by other V2X technologies, for example by the infrastructure, i.e., by V2I communication. For this reason, in this study, we investigate the exposure levels of road users in a realistic urban vehicular scenario that includes not only V2V but also V2I communication technologies. For that purpose, we considered a real 3D map of the city center of Manhattan as a realistic vehicular urban scenario. The electric field (E-field) due to V2V and V2I antennas was computed with the raytracing deterministic method and then used to assess the dose absorbed by road users in the urban scenario specifically generated by V2V and V2I in a far-field exposure condition. We computed the dose absorbed by a generic road user in such a realistic urban scenario by varying the distance between this generic road user and the radiofrequency (RF) sources in the scenario, i.e., V2V and V2I antennas, and by considering road users of different anatomical characteristics, i.e., with different Body Mass Index (BMI) and size.

## 2. Materials and methods

Figure 1 illustrates a schematic depiction of the steps followed to compute the dose of RF fields absorbed by the generic road user due to the RF source in the urban scenario.

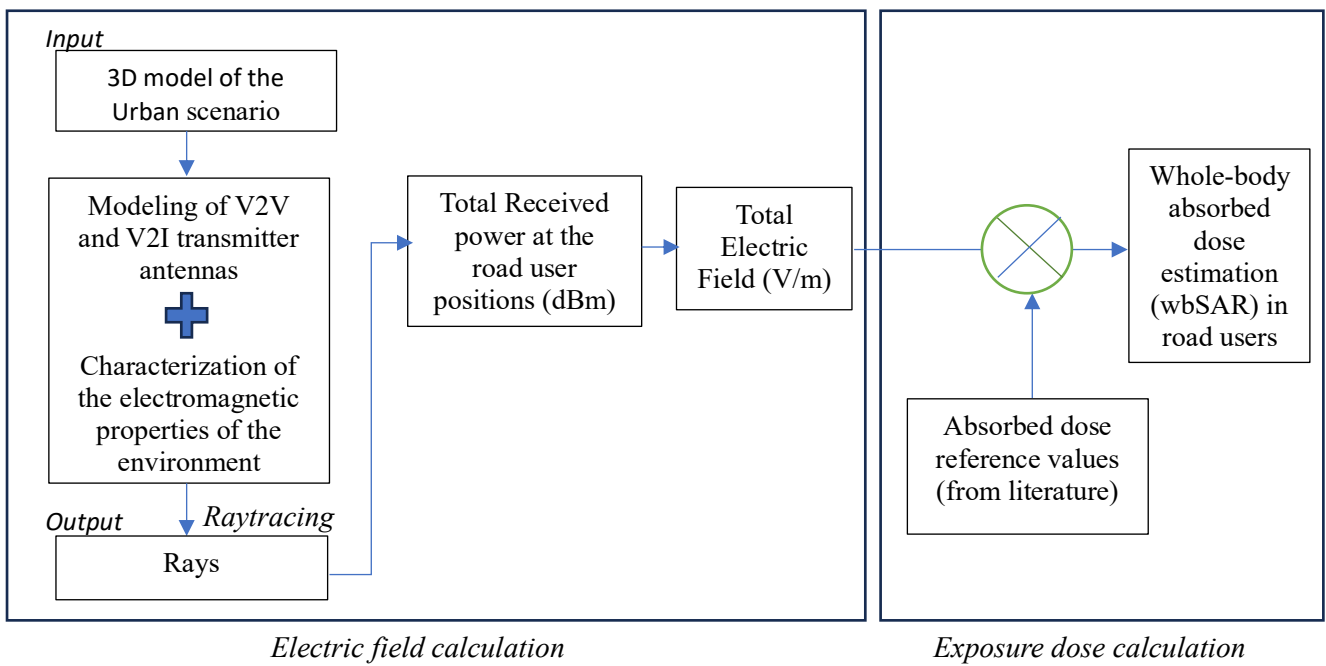

Figure 1. Schematic view of the deterministic approach used to compute the dose absorbed by a generic road user in the exposure scenario considered in this study.

### 2.1 Urban Scenario

To conduct a realistic assessment of environmental exposure levels, we used the 3D map of central Manhattan as available in Remcom's Wireless Insite tool [24]. We analyzed a portion of total dimension 85.5 m x 90 m of the original and larger map of central Manhattan; the analyzed area comprises one road intersection and includes a multitude of features that typically characterize a realistic urban environment, namely road terrain (made of asphalt), buildings of varying size and height (maximum height reached of about 80 m), wet earth (grass), trees, and a total of five vehicles.

## 2.2 Setup of the exposure scenario

In this section, we refer to the term *exposure scenario* as a specific set of conditions in the 3D urban map presented in the previous section 2.1, under which the RF dose absorbed by human models will be investigated. The electromagnetic sources in the exposure scenario consist of V2V and V2I antennas for vehicular communication. In the pursuit of a complete investigation of the dose absorbed by road users, we focused on three different exposure scenarios of increasing complexity and representative of three daily life situations. The three exposure scenarios investigated are shown in Figure 2. Scenario 1 (Figure 2A) comprises only one transmitting vehicle (the blue one). Scenario 2 (Figure 2B), in addition to the transmitting car of scenario 1, comprises a RSU transmitting antenna that was positioned adjacent to a building facade to replicate its intended placement on a traffic light. Scenario 2 will help us to understand the contribution of the EMF field emitted by the RSU on the dose absorbed by the generic road user. Finally, in scenario 3 (Figure 2C), in addition to the transmitters used in scenario 2, we introduced an additional four transmitting vehicles for a total of five transmitting vehicles to see how an increased number of vehicular antennas would affect the dose absorbed by the road user. This condition represents the worst-case exposure scenario.

All the vehicles in the scenarios act like static objects (i.e., they are not moving). In scenario 3 we positioned the additional four vehicles in the vicinity of the original transmitting vehicle of scenario 1 at random distances, within the distance $d_{lim}$ which we defined as the distance within which the dose of exposure induced by the transmitting car in scenario 1 was higher than 70% of the $99^{th}$ percentile of the maximum exposure dose in the analyzed urban area. We assessed the exposure dose by calculating the whole-body SAR - wbSAR – which is defined as the ratio between the power of the RF field absorbed by the body and the total mass of the body. The next Section 3.1 details how much this $d_{lim}$ value was.

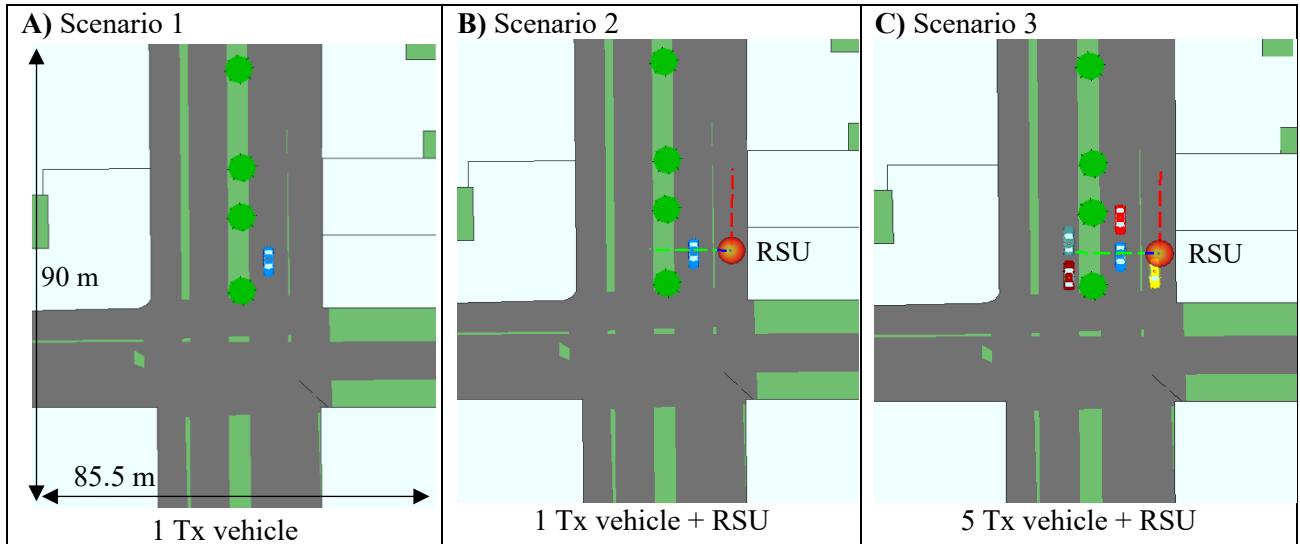

Figure 2. Illustration of the 3D urban area top view of dimension 85.5 m x 90 m with the 3-exposure scenarios. **A)** Scenario 1 consists of only one vehicle transmitting; **B)** In scenario 2 in addition to the vehicle transmitting also the RSU is activated; **C)** Scenario 3 is the same of scenario 2 but with 4 more transmitting vehicles surrounding the blue one.

## 2.3 Modeling of V2V and V2I antennas and electromagnetic characterization of the environment

In our scenarios, we considered two different types of RF transmitters (Tx): the V2V antennas and the RSU antenna. V2V and RSU antennas were modeled as omnidirectional antennas, i.e., as half-wave dipoles [16][17][18][21][25][26] operating at 5.9 GHz, with a bandwidth of 10 MHz [7][27]. For all antennas, the input power was set to 33 dBm (which is the maximum allowable power in the EU [7]), with a gain of 0 dBi. Figure 3 shows the location and radiation pattern of each transmitting antenna. The V2V antennas were mounted on the roof of the vehicles (one antenna per vehicle) at a height of 1.7 m from the ground [28][29], while the RSU antenna was placed next to the building's façade, at 5 m from the ground and tilted 10 degrees toward the ground according to 3GPP recommendations [27].

To assess the variability of road user exposure with the position from the transmitting antennas, we computed the E-field generated by the transmitting antennas in the analyzed urban area (Figure 2) on an evaluation grid of regularly spaced points on the *xy* plane. The points on the evaluation grid were spaced by 3 m and were modeled as generic receivers (Rx), i.e., as isotropic antennas. The Rx grid was placed at different heights along the *z-axis* (as detailed in Section 2.5).

The dielectric properties of the objects included in the urban area (Figure 2) were set according to the ITU database [30] and literature data [31] (Table 1).

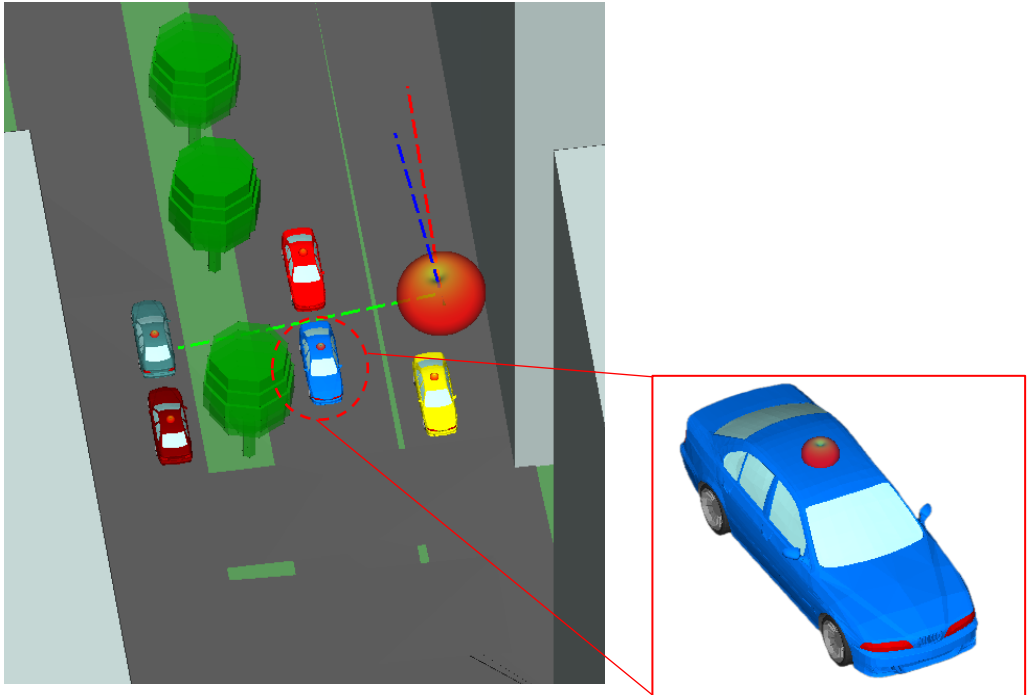

Figure 3. Illustration of the positions of the transmitting antennas and the corresponding radiation patterns in scenario 3. This scenario comprises 1 V2V antenna for each vehicle and 1 RSU next to the building's facade.

Table 1. Dielectric properties of the objects modeled in the analyzed urban area at the frequency of 5.9 GHz. DHS: Dielectric half-space, OLD: one-layer dielectric, PEC: Perfect electric conductor.

| Objects | Materials | Type | Conductivity (S/m) | Permittivity (adimensional) |
|---|---|---|---|---|
| Terrain | Wet earth | DHS | 1.215 | 15.76 |
| Pavement | Asphalt | DHS | 0 | 5.72 |
| Buildings | Concrete | OLD | 0.12 | 5.31 |
| Leaf (tree) | Leaf | Biophysical | 0.39 | 26 |
| Branches (tree) | Wood | Biophysical | 0.39 | 20 |
| Hub caps (vehicle) | Plastic | OLD | 0.05 | 4.70 |
| Lights (vehicle) | Plastic | OLD | 0.05 | 4.70 |
| Mirrors (vehicle) | Glass | OLD | 0.23 | 6.27 |
| Bumper (vehicle) | Plastic | OLD | 0.05 | 4.70 |
| Tires (vehicle) | Rubber | OLD | 0.06 | 2.6 |
| Body vehicle | Metal | PEC | / | / |
| Undercarriage (vehicle) | Metal | PEC | / | / |

## *2.4 Electric field calculation through raytracing*

To calculate the E-field generated by V2V and V2I antennas, we used raytracing, as provided by Remcom's Wireless Insite EMF propagation tool [24]. We used the X3D propagation model implemented in the Wireless Insite tool [24]. This propagation model is the most recent model for outdoor investigation that effectively merges two ray tracing methodologies, i.e., the Shooting and Bounding Rays (SBR) method and the Image theory (IM) method. More precisely, it combines the best of these two methodologies, implementing first the

SBR method to launch rays uniformly across a spherical surface centered at the transmitter with a specific *ray spacing*; secondly, the IM method is used to perform an Exact Path Calculator (EPC) correction to determine the precise rays between each Rx point [24]. All the parameters used for the raytracing simulations are shown in Table 2. As described in [24], these parameters allow us to accurately predict electromagnetic wave propagation in the environment considered in our study.

To better characterize the E-field, we considered in our raytracing simulations the Diffuse Scattering (DS) phenomenon, alongside reflection, refraction, and diffraction effect (Table 2). This way the E-field can be better characterized in the Non-Line-of-Sight (NLOS), as it was demonstrated by [32] that the DS is mostly relevant in NLOS and negligible in Line-of-Sight (LOS). More precisely, in this study, the *Directive model* was used to assess the contribution of the DS. This choice was motivated by the fact that, when compared to the other DS models used in literature (i.e., *Lambertian model*), Vittorio degli Espositi et al., [33] demonstrated that the *Directive model* best follows the experimental measurement (also demonstrated by [34]). Specifically, the DS was applied to the building walls; as parameters were set the *scattering factor* "S" = 0.45, the *cross-polarization fraction* "K-pol" = 0.4, and the *alpha* value, i.e., the amplitude of the lobe scatterer ray, as 4 [33][34][35][36].

Furthermore, to predict the path loss due to foliage and vegetation the Weissberger's model was used [37].

Table 2. Computational parameters of the raytracing simulations.

| Propagation model: X3D |
|---|
| Ray spacing: 0.2° [24] |
| Number of reflections: 6 [24] |
| Number of refractions: 0 [24] |
| Number of diffractions: 1 [24] |
| Received threshold: -250 dBm |
| DS model: Directive model [33] |
| *Scattering factor* (S): 0.45 |
| *Cross-polarized factor* (K-pol): 0.4 |
| *Alpha*: 4 |
| Vegetation/foliage model: Weissberger's model [37] |

## *2.5 Human models investigated*

To assess the RF dose absorbed by the generic road user, we considered three human models of different body sizes and ages. This choice allowed us to investigate the variability of exposure levels due to the different anatomical characteristics of the road user. We selected from the Virtual Family Population (ViP) (https://itis.swiss/virtual-population/virtual-population/overview/) of human models two adults, i.e., one male model - 'Duke' - and one female model - 'Ella' - and a female child, called 'Nina'. Table 3 reports the different anatomical characteristics of the three human models. To assess the dose of RF absorbed at the most crucial part of the body – the head - we calculated the E-field at a height along the z-axis that corresponds to the level of the head of each of the three human models, that is at $z = 1.7$ m (which corresponds the head level of the model 'Duke', $z = 1.5$ m (head level of 'Ella'), and $z = 0.85$ m (head level of 'Nina').

Table 3. Anatomical characteristics of the three human models investigated in this study.

|  | Age (years) | Sex | Height (m) | Weight (kg) | BMI (kg/m$^2$) |
|---|---|---|---|---|---|
| Duke | 34 | male | 1.77 | 70.2 | 22.4 |
| Ella | 26 | female | 1.63 | 57.3 | 21.6 |
| Nina | 3 | female | 0.92 | 13.9 | 16.4 |

## *2.6 Evaluation of the absorbed dose of RF fields*

The dose of RF fields absorbed by the generic road user (here represented by the three different human models) was assessed by calculating the Specific Absorption Rate (SAR) over the whole body (wbSAR [W/kg]).

In far-field exposure conditions, the wbSAR of the generic road user of body mass index $BMI_{ru}$ [kg/m$^2$], can be obtained by this formula [38]:

$$wbSAR = (E_{inc}/E_{ref})^2 \cdot (BMI_{ref}/BMI_{ru}) \cdot SAR_{ref} \qquad (1)$$

where $E_{inc}$ is the incident E-field obtained from the Raytracing simulation (expressed as the root mean square value of the E-field in V/m) at the position of the Rx on the evaluation grid and $E_{ref}$ is the reference incident E-field (V/m) that was used to calculate the reference $SAR_{ref}$ (W/kg) in a reference human body of body mass index $BMI_{ref}$ (kg/m$^2$). $E_{ref}$ is equal to 2.45 V/m [38].

Because the human models investigated here are the same as those considered as 'reference' by the authors in [38] the values of $BMI_{ref}$ and $BMI_{ru}$ are the same; therefore, in the formula (1) their ratio is equal to 1.

Table 4 reports the reference $SAR_{ref}$ values as calculated in [38].

Table 4. $SAR_{ref}$ values of the three human models investigated in this study. The $SAR_{ref}$ values obtained by [38] were calculated in far-field condition at 5.8 GHz for an incident E-field of 2.45 V/m.

| Human models | $SAR_{ref}$[24] (W/kg) |
|---|---|
| Duke | $3.6 \cdot 10^{-5}$ |
| Ella | $4.0 \cdot 10^{-5}$ |
| Nina | $6.0 \cdot 10^{-6}$ |

It must be noted that Liorni et al., [38] calculated the $SAR_{ref}$ values at a frequency of 5.8 GHz. The antennas used in vehicular connectivity, based on the protocol of IEEE 802.11p [7], operate at a nominal frequency of 5.9 GHz in the frequency band 5.855-5.925 GHz [7]. While this frequency for the vehicular scenario deviates slightly from 5.8 GHz, an examination of human tissue dielectric properties [39][40] (relevant for absorbed dose calculation), reveals negligible differences between 5.8 GHz and 5.9 GHz. Specifically, on average, the conductivity at 5.8 GHz is 0.98 times that at 5.9 GHz, and the relative permittivity at 5.8 GHz is the same as that at 5.9 GHz [39][40]. Given these minimal differences in dielectric properties at these frequencies, it is reasonable to consider the $SAR_{ref}$ obtained at 5.8 GHz [38] as a reliable approximation for $SAR_{ref}$ values at 5.9 GHz.

*2.7 Statistical parameters investigated*

We calculated the median, maximum, 25[th] percentile, 75[th] percentile, 99[th] percentile, and the skewness of wbSAR for each human model, in each of the three exposure scenarios. We then compared the maximum wbSAR to the basic restriction limits of exposure in the 100 kHz – 300 GHz range set in the ICNIRP [22] and IEEE [23] guidelines.

# 3. Results

## *3.1 Calculation of $d_{lim}$ and $d_i$ distances*

*$d_{lim}$ calculation* - As explained in previous Section 2.2, the vehicles in the scenario were placed around the blue vehicle within the distance "$d_{lim}$", which is defined as the distance within which the wbSAR induced by the E-field generated by the blue car in Scenario 1 was at least 70% of the 99[th] percentile of wbSAR distribution. The level of 70% of the 99[th] percentile of the wbSAR represents the fraction of absorbed energy which corresponds to -3 dB of the maximum.

Figure 4A shows as a practical example the steps we followed to calculate $d_{lim}$. First, we computed the total received power (dBm) generated by the blue transmitting vehicle of scenario 1, as evaluated over the grid of Rx at the height $z = 1.5$ m (Figure 4A). Then (Figure 4B) we calculated the E-field derived from the total received power and (Figure 4C) the wbSAR (calculated from (1) using the E-field) for the human model 'Ella'. The E-field and the wbSAR were plotted in Figure 4 as functions of the distance from the blue transmitting vehicle. Finally (Figure 4C), we found the distance $d_{lim}$ from the blue car (represented in Figure 4C by the blue dashed line) for which the wbSAR of this human model was at least 70% of the 99[th] percentile of the wbSAR distribution. For this human model $d_{lim}$ was calculated to be equal to 8 m.

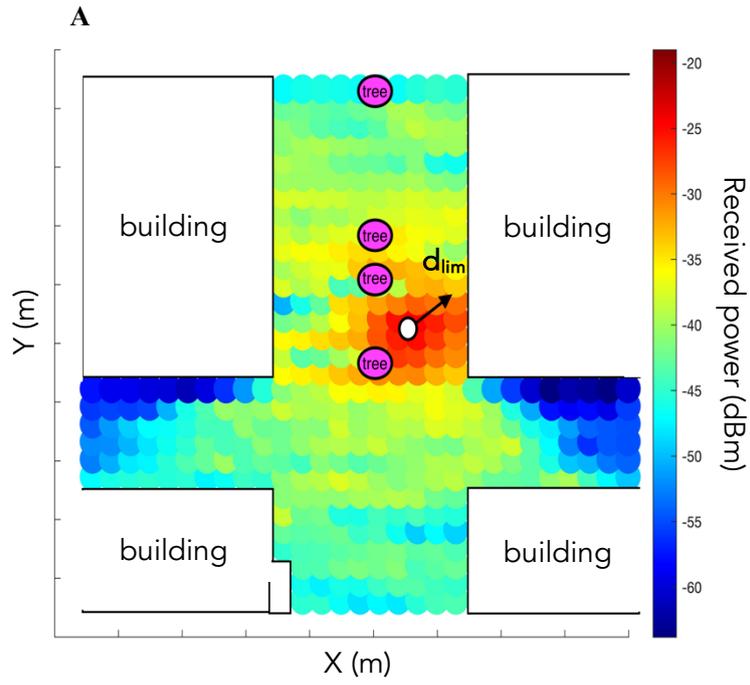

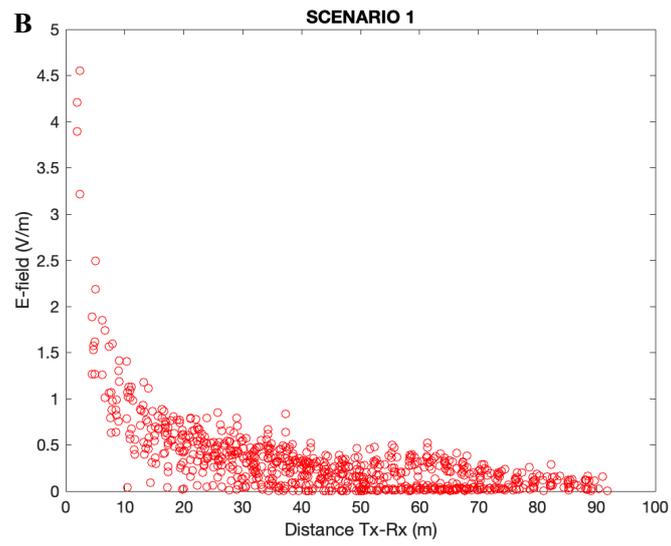

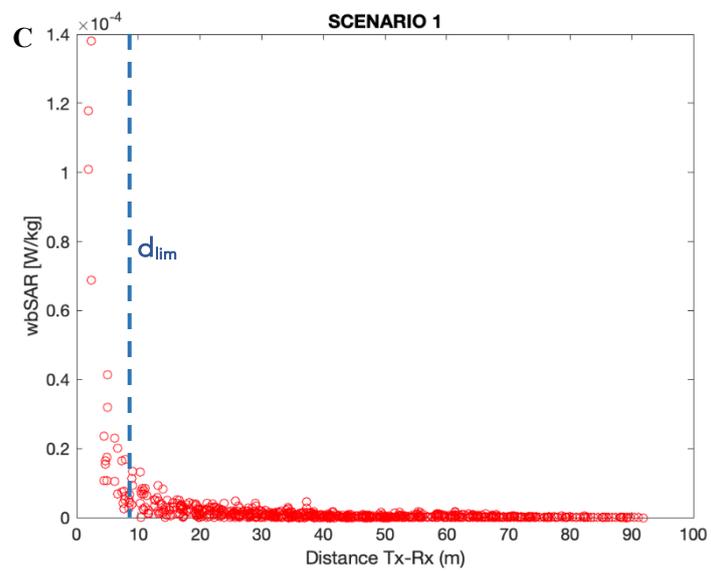

Figure 4. **A)** Color map of the total received power (dBm) investigated at $z$ =1.5 m. The blue transmitting vehicle is represented by the white dot. For a better clarification is also represented the distance $d_{lim}$ from the blue transmitting vehicle. The locations of the trees are represented with magenta dots. **B)** Corresponding E-field values (V/m) obtained as a function of the distance from the transmitting vehicle (the white dot). **C)** wbSAR values (W/kg) obtained from (1) by using the E-field displayed in panel B for the adult model 'Ella' as a function of the distance from the transmitting vehicle. The blue dashed line represented the corresponding $d_{lim}$ calculated as the distance within with the E-field (panel B) induced wbSAR values higher than the 70% of the 99$^{th}$ of the wbSAR distribution.

The distance $d_{lim}$ within which the wbSAR value was at least equal to 70% of the 99$^{th}$ percentile was found to range from 6.6 m to 10.6 m across the human models. To ease the comparison of the absorbed dose across the different human models in the different scenarios, we decided to perform all the analyses by considering the same $d_{lim}$ value of 10.6 m for all the human models. This $d_{lim}$ value also takes into account the constraints imposed by the geometry of the environment (such as the size of the vehicle and the minimal distance between the vehicles). This choice of $d_{lim}$ means that in scenario 3 where we considered the presence of more than a single transmitting car, all the vehicles were placed within 10.6 m of the blue vehicle.

*$d_i$ calculation* – E-field and wbSAR values in the following sections are calculated and displayed only for the points on the evaluation grid (described in Section 2.3) that fall within a so-called 'region of interest' (ROI) that we defined as the region that contains all the areas of influence of radius $d_{lim}$ of the various transmitting vehicles as shown in Figure 5. Figure 5 shows the region of interest ROI represented by a black square. Furthermore, as we took as the origin of the reference system the blue transmitting vehicle (presented in all three different scenarios), the analysis of E-field and wbSAR values within the ROI is defined as a function of the distance, called "$d_i$", from the blue transmitting vehicle. The distance $d_i$ is illustrated in Figure 5 with a black arrow from the blue Tx car (reference system) within the ROI.

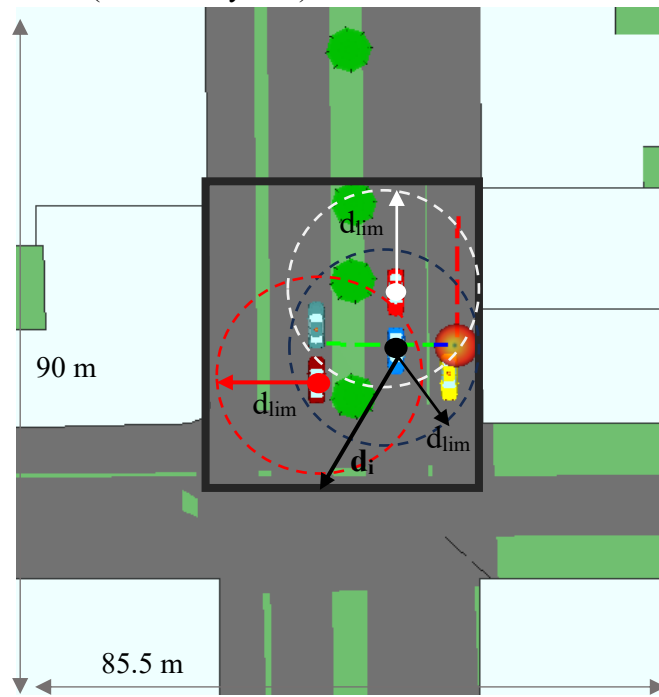

Figure 5: Depiction of the region of interest ROI represented by a black square. ROI includes the region of influence of all the transmitting vehicles defined as $d_{lim}$ that is the distance beyond which the wbSAR values are higher than 70% of the 99$^{th}$ percentile. The dashed circles represent the area of influence equal to $d_{lim}$, for each car. Furthermore, within the ROI, the E-field and wbSAR values in each scenario are evaluated as a function of the distance $d_i$ from the blue car (reference system). For the sake of clarity, we showed in the figure the area of influence for only three of the five vehicles considered in our study.

### *3.1 E-fields as a function of the distance $d_i$*
Figure 6 shows an example of the E-Field (V/m) evaluated within the region of interest ROI at the height $z$ =1.5 m for each of the three exposure scenarios as a function of the distance $d_i$ from the blue vehicle (Figure 5).

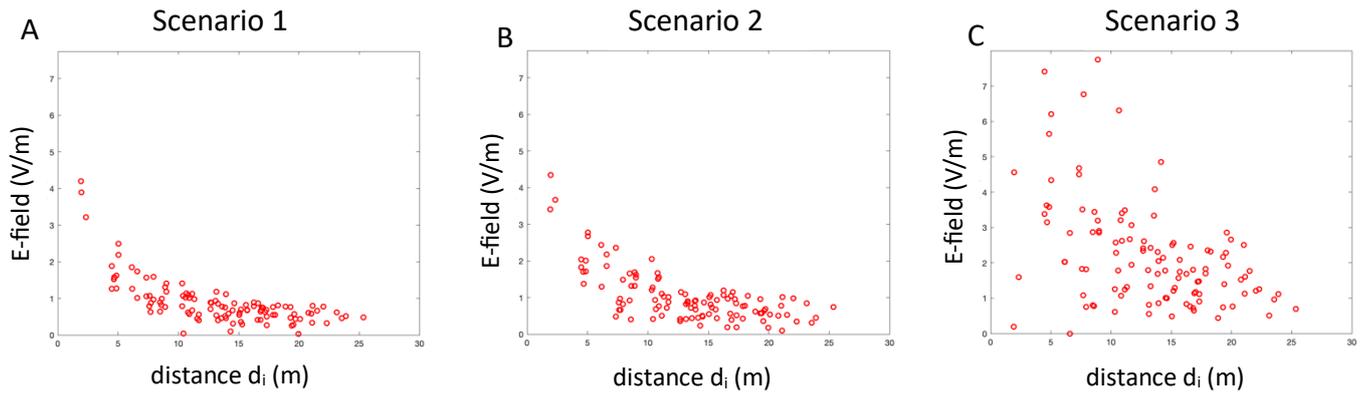

Figure 6. E-field (V/m) values computed at the height z =1.5 m as a function of the distance $d_i$ from the blue vehicle in the exposure scenario 1 (only 1 transmitting vehicle) **(A)**, scenario 2 (one transmitting vehicle + one transmitting RSU) **(B)**, and scenario 3 (five transmitting vehicles + one transmitting RSU) **(C)**. The E-field is represented from the minimum up to the 99$^{th}$ percentile value. Please note that the values shown in panel A are the same as those already displayed in Figure 4C, except that here only the values within the ROI are shown.

From Figure 6, it is evident that the E-field at any given distance from the blue car is not a unique value but exhibits some variability. This was because the E-field at any given distance depends on the different propagation conditions encountered along the optical trajectory between the blue car and the point at which the E-field was calculated.

As expected, as the number of antennas increased, the 99$^{th}$ percentile E-field values increased, ranging from 4.3 V/m in the scenario 1 (Figure 6A) to 8.5 V/m in scenario 3 (Figure 6C) where there were the maximum number of transmitting vehicles.

In scenarios 1 and 2 the maximum of the E-field was found near the vehicle (at 2 m); the 99$^{th}$ percentile slightly increased from 4.3 V/m in scenario 1 to 4.7 V/m in scenario 2. This would mean that the additional effect that RSU in scenario 2 would have on the exposure level generated only by the transmitting car was almost negligible. In scenario 3, where more than one vehicle is transmitting, we could see that the E-field was generally significantly higher than in scenario 1 and 2 because of the summation effect of the fields generated by multiple transmitting cars.

Figure 6 refers specifically to the E-field computed at z= 1.5 m. Regarding the other heights investigated, i.e., $z$ =1.7 m and $z$= 0.85 m, the E-field generally showed a similar behavior to that one obtained at z=1.5 m. Differences were seen in the maximum values and, to a lesser extent, their locations. At z= 1.7 m, the E-field was calculated at the same height as the antennas mounted on the cars; as such, the E-field obtained was higher than that obtained at z =1.5 m and z= 0.85 m, with a 99$^{th}$ percentile of 5.6 V/m in scenario 1 and 9 V/m in scenario 3. The maximum E-field value was located at a slightly farther distance than that observed at z=1.5m, namely at 2.3 m in scenarios 1 and 2 and at 10 m in scenario 3. On the contrary, the E-field values at z =0.85 m were the lowest among the three heights investigated (with the 99$^{th}$ percentile ranging from 1.9 V/m in scenario 1 to 3.6 V/m in scenario 3), because the points at which E-field was evaluated were farther from the V2V antennas. Furthermore, at z=0.85 m, among the three scenarios, the maximum E-field was located at 6.5 m in scenarios 1 and 2 and at 8 m in scenario 3. The maximum E-values were located far from the vehicle mostly because the roof of the car shields the propagation of the radiated field downward.

### *3.2 Whole-Body Specific Absorption Rate*
Figure 7 shows the boxplot of the wbSAR distributions for each human model across the different exposure scenarios here investigated. As expected, it can be seen from Figure 7 that as the number of antennas increased (from scenario 1 to 3) the maximum wbSAR value increased, reaching the highest values in scenario 3. The highest wbSAR was found in 'Duke' (4.9·10$^{-4}$ W/kg), followed by 'Ella' (3.8·10$^{-4}$ W/kg), and 'Nina' (0.13·10$^{-4}$ W/kg). The differences in the wbSAR values among the human models were mainly attributed to (*i*) variation in the incident E-field at the three different heights of the heads and (*ii*) the different scaling factor SAR$_{ref}$ that was used in (1) to calculate the absorbed dose. More precisely, the child model 'Nina' exhibited wbSAR values that were generally one order of magnitude lower than those of the adult models. This is due to the SAR$_{ref}$ value of the child model being one order of magnitude lower than that of the adult models (Table 4). Besides the different SAR$_{ref}$, the wbSAR values for the child were expected to be lower compared to that of the adult models because the E-field reaching the child model was lower. As commented above, the E-field at the height of the child head was significantly lower than that observed at the height of the head of the two adult models. Because the SAR$_{ref}$ values (Table 4) and the E-field computed at the head models of the two adults were similar, the wbSAR values of the two adults ('Duke' and 'Ella') showed very similar wbSAR values (Figure 7).

It is noteworthy to observe from Figure 7 that scenarios 1 and 2 show nearly the same exposure levels. The only difference between these two scenarios was the activation of the RSU in scenario 2 (Figures 2A and 2B). More precisely, the median of the exposure levels in scenario 2 was, on average, only 1.3 – 1.6 times those obtained in scenario 1. This would mean that the additional effect of RSU on the exposure on road user human models was minimal, and as such can be considered negligible compared to those generated by the V2V antennas. This was mainly because the RSU was at a higher height (5 m above the ground) compared to the antenna placed on the vehicle.

Table 5 reports for each human model and each exposure scenario the median, 25$^{th}$, 75$^{th}$, and 99$^{th}$ percentiles (also reported in Figure 7), and skewness. The median wbSAR (Table 5) ranged from 3.8·10$^{-7}$ W/kg to 2.4·10$^{-5}$ W/kg across the human models and scenarios and was generally two orders of magnitude lower than the 99$^{th}$ percentile. Finally, it was observed that the distribution of wbSAR for all the human models and all exposure scenarios had a strong positive skewness. This means that most of the wbSAR values were distributed in the interval with low exposure levels.

All the wbSAR values obtained from the present study were well below the basic restriction limits of exposure in the 100 kHz – 300 GHz range as recommended by the ICNIRP and IEEE of 0.08 W/kg [22][23].

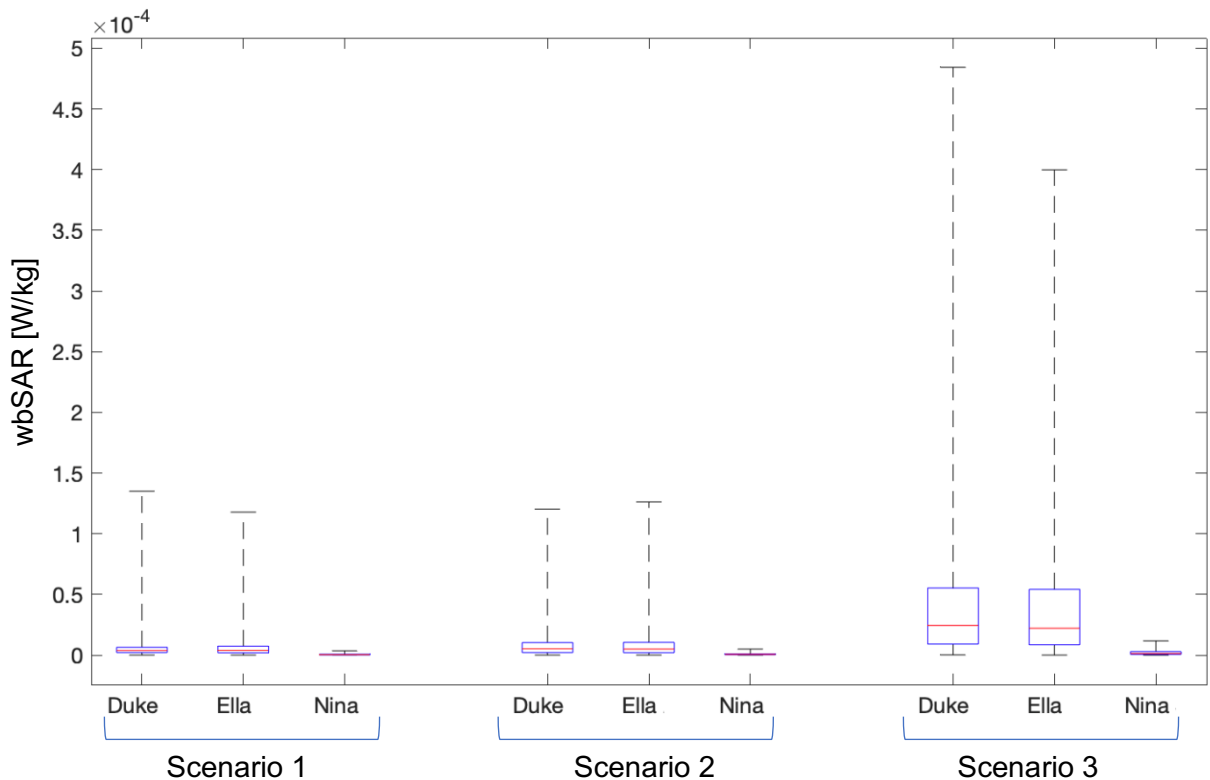

Figure 7. Boxplot of the wbSAR [W/kg] distributions among all the human models and scenarios investigated. The red line represents the median value, the extreme of the box the 25$^{th}$ and 75$^{th}$ percentile, and the lower and upper whisker the 1$^{st}$ and 99$^{th}$ percentile, respectively.

Table 5. Statistical parameters of the wbSAR distribution on the three human models for each exposure scenario investigated. The statistical wbSAR values are the same illustrated in Figure 7. All the parameters except the skewness are expressed in W/kg.

|  | Scenario 1 (1 Tx car) | Scenario 2 (1 Tx car + 1 RSU) | Scenario 3 (5 Tx car + 1RSU) |
|---|---|---|---|
| Duke (h=1.7 m) | $25^{th} - 75^{th}$ perc= $2.1 \cdot 10^{-6} - 6.6 \cdot 10^{-6}$<br>$99^{th}$ perc = $2 \cdot 10^{-4}$<br>Median= $3.7 \cdot 10^{-6}$<br>Skewness= 7.2 | $25^{th} - 75^{th}$ perc = $2.1 \cdot 10^{-6} - 1 \cdot 10^{-5}$<br>$99^{th}$ perc = $1.6 \cdot 10^{-4}$<br>Median= $5.2 \cdot 10^{-6}$<br>Skewness= 5.3 | $25^{th} - 75^{th}$ perc = $9 \cdot 10^{-6} - 5.6 \cdot 10^{-5}$<br>$99^{th}$ perc = $4.9 \cdot 10^{-4}$<br>Median= $2.4 \cdot 10^{-5}$<br>Skewness= 2.4 |
| Ella (h=1.5 m) | $25^{th} - 75^{th}$ perc= $2 \cdot 10^{-6} - 7.5 \cdot 10^{-6}$<br>$99^{th}$ perc = $1.2 \cdot 10^{-4}$<br>Median= $3.8 \cdot 10^{-6}$<br>Skewness = 4.7 | $25^{th} - 75^{th}$ perc = $2 \cdot 10^{-6} - 1 \cdot 10^{-5}$<br>$99^{th}$ perc = $1.5 \cdot 10^{-4}$<br>Median= $5 \cdot 10^{-6}$<br>Skewness= 4.8 | $25^{th} - 75^{th}$ perc = $8.5 \cdot 10^{-6} - 5 \cdot 10^{-5}$<br>$99^{th}$ perc = $4.8 \cdot 10^{-4}$<br>Median= $2.2 \cdot 10^{-5}$<br>Skewness= 3.8 |
| Nina (h=0.85 m) | $25^{th} - 75^{th}$ perc= $1 \cdot 10^{-7} - 7.6 \cdot 10^{-7}$<br>$99^{th}$ perc = $0.4 \cdot 10^{-5}$<br>Median= $3.8 \cdot 10^{-7}$<br>Skewness= 2.6 | $25^{th} - 75^{th}$ perc = $2.9 \cdot 10^{-7} - 1 \cdot 10^{-6}$<br>$99^{th}$ perc = $0.5 \cdot 10^{-5}$<br>Median= $6.2 \cdot 10^{-7}$<br>Skewness= 3.2 | $25^{th} - 75^{th}$ perc = $6.7 \cdot 10^{-7} - 3 \cdot 10^{-6}$<br>$99^{th}$ perc = $1.3 \cdot 10^{-5}$<br>Median= $1.3 \cdot 10^{-6}$<br>Skewness= 2.4 |

Figure 8 shows as an example of the spatial distribution within the ROI, the wbSAR in the adult 'Duke', obtained in each of the three exposure scenarios; in all panels of figure 8 the wbSAR values were related to the maximum $99^{th}$ percentile obtained among the human models and scenarios investigated, i.e. $4.9 \cdot 10^{-4}$ W/kg in 'Duke' in scenario 3 (Table 5).

Because the SAR was strictly related to the E-field (formula 1), in scenarios 1 and 2 (Figure 8A and 8B, respectively) the wbSAR values followed the same exponential decade as the corresponding E-field (Figures 6A and 6B). Considering the blue Tx vehicle as a reference system, the maximum in adults (Duke and Ella) was observed close to the vehicle at $d_i = 2$ m. On the contrary, 'Nina', with its low height did not receive absorbed dose near the vehicle as the car roof shielded most of the radiation downward, resulting in its maximum of $0.5 \cdot 10^{-5}$ W/kg located at $d_i = 6.6$ m from the vehicle.

Conversely, in scenario 3 (Figure 8C) that involved more than one transmitting vehicle, the spatial distribution of E-field was different from the one observed in scenarios 1 and 2. It is possible to observe that the E-field exhibited, instead of a single peak, multiple peaks with high absorbed doses distributed in the investigated area (ROI) (following the same trend as the E-field in Figure 6C). 'Duke' and 'Ella' exhibited peaks of absorbed dose up to about 11 m from the blue Tx vehicle, while 'Nina' up to 14 m. Precisely, among these multiple peaks, the maximum for 'Duke' and 'Ella' was at 10.5 m and 9 m from the blue car, respectively, whereas 'Nina' was at 8 m from the blue car. These maximum values were located at distances at which there was the influence of two or more Tx vehicles.

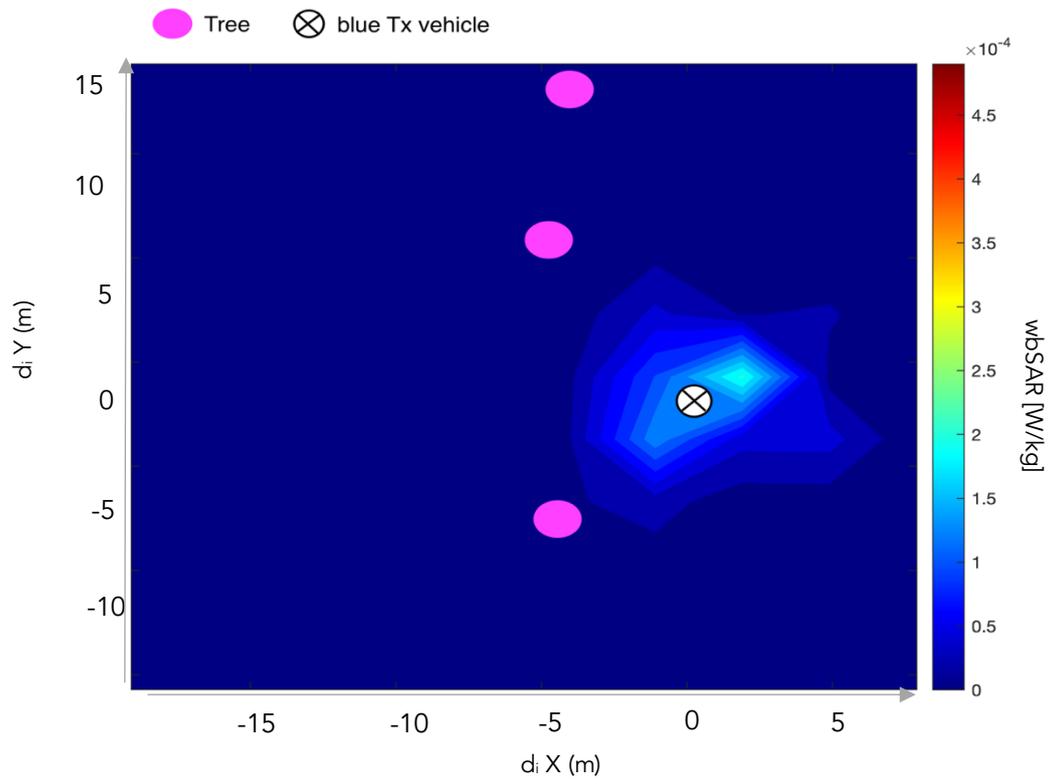
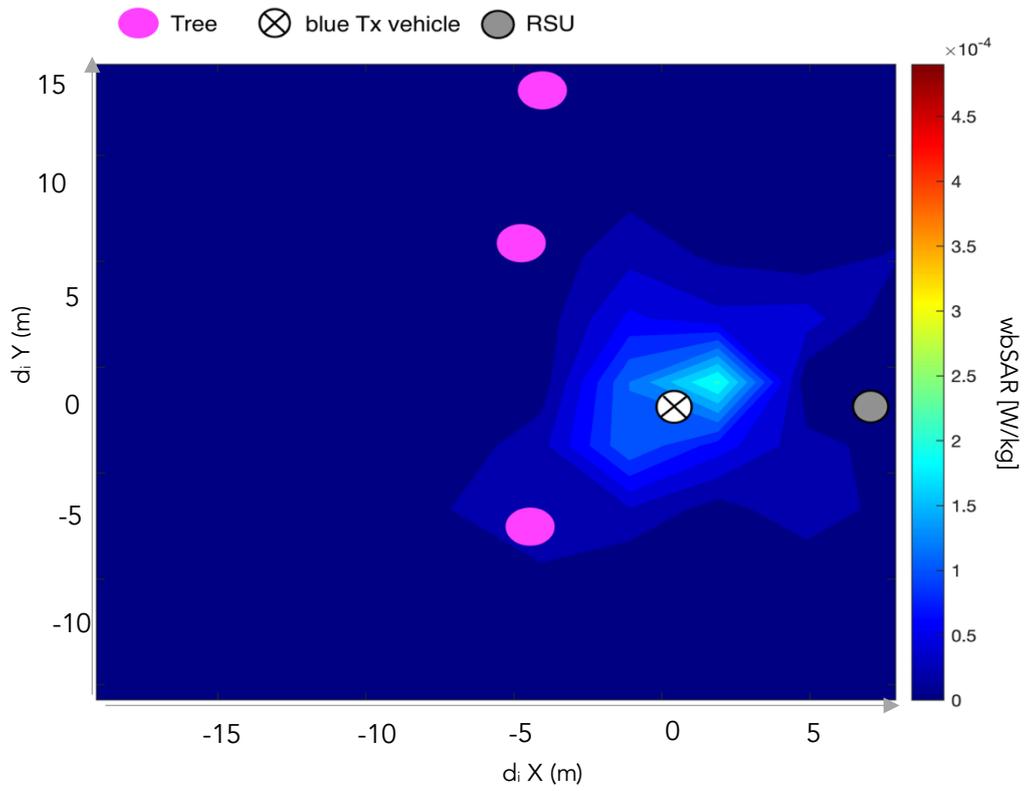

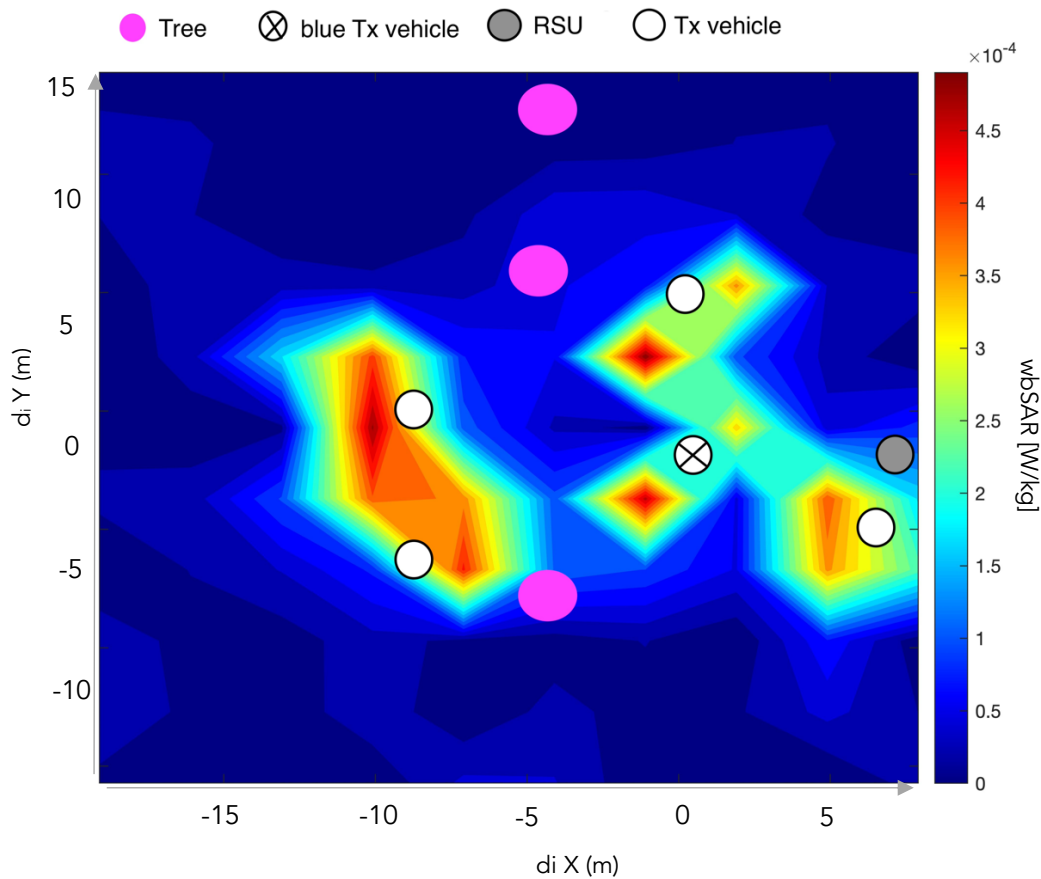

Figure 8. Color map of the wbSAR (W/kg) values of 'Duke' human models in exposure scenario 1 **(A)**, scenario 2 **(B)**, and scenario 3 **(C)** within the region of interest ROI. The origin of the *x*- and *y*-axis, which delimit the region of interest ROI, is centered on the blue Tx vehicle which is the reference system. These values were calculated with an incident E-field computed at a height of *z* =1.7 m.

## 4. Discussion

In this study, we investigated for the first time the RF exposure levels simultaneously generated by different V2X communication technologies on a road user in an urban scenario. More precisely, we investigated the wbSAR values due to the concurrent presence of V2V and V2I communication technologies operating both at the nominal frequency of 5.9 GHz [7]. The urban scenario analyzed in this study corresponded to a portion of the real map of Manhattan city and comprised many geometrical features typically seen in realistic urban scenarios. We combined the use of a deterministic approach, i.e., *Raytracing* with an analytical formula taken from literature [38] to obtain the dose absorbed under far-field exposure conditions. We assessed the dose absorbed in three different human models (two adults + one child) of different anatomical characteristics by computing the wbSAR that was induced by the E-field generated by the V2V and V2I transmitters. As a result of this methodology, we were able to investigate the wbSAR variability as a function of *i)* the distance from the transmitting antenna(s), and *ii)* different anatomical characteristics of the human models. We explored how much was the dose absorbed by the human models in three different exposure conditions of increasing complexity.

As a general observation, with only one transmitting vehicle (scenario 1) the exposure levels were the lowest among the three scenarios, with the 99$^{th}$ percentile wbSAR values in a range of $3.5 \cdot 10^{-6}$ W/kg - $2 \cdot 10^{-4}$ W/kg across the human models. On the contrary, when multiple Tx vehicles and the RSU were switched on (scenario 3), the exposure levels were the highest reaching a range of $1.3 \cdot 10^{-5}$ W/kg – $4.9 \cdot 10^{-4}$ W/kg. Among the human models, the adults exhibited always the highest wbSAR values (up to $4.9 \cdot 10^{-4}$ W/kg), while the child model was always the lowest.

We also found that, depending on the number of V2V antennas activated, the human models were affected by radiation from 2 m up to 11 m from the transmitting vehicle. The effect of the RSU antenna on the exposure level of the road user was negligible.

Our analysis demonstrated that exposure levels on different human models, when the V2V antennas were placed on the roof of the car, were predominantly affected by body size, in particular by height. Indeed, among the two adults (Duke and Ella), of similar heights, the exposure levels were almost the same. On the contrary, the wbSAR of the child, because of its lower height, significantly differed from those of adults. The child model had always wbSAR values that were one order of magnitude lower than those of adults. These differences in the wbSAR values of the human models were mostly due to *i*) the scale factor 'SAR$_{ref}$' used in the analytical formula (1) for computing the wbSAR values, and *ii*) the E-field values in the ROI, which highly depend on the geometrical features and characteristics of the urban scenario. The child, with a low height of 0.92 m, when placed close to the vehicle was less exposed to radiation due to the higher distance from the transmitting antennas (positioned at 1.7 m) and the shielding effect of the car roof which blocked most of the radiation downwards. On the contrary, the adults with a height more comparable to that of the V2V antenna, i.e., 1.6 m for 'Ella' and 1.77 m for 'Duke', were more exposed when they were placed close to the Tx vehicle. More precisely, in scenarios 1 and 2 the adults received the highest dose absorbed at 2 m from the Tx vehicle, while the child at 6.6 m far from the Tx vehicle. In scenario 3, with multiple Tx vehicles, the maximum dose absorbed resulted in around 10 m for adults and 8 m for the child.

It is important to highlight that all the wbSAR values found in the current study were well below the limits imposed by the ICNIRP [22] and IEEE [23] guidelines, which set a limit of 0.08 W/kg for whole-body exposure over an averaging interval of 30 minutes.

To the best of our knowledge, this is the first study that assessed the exposure levels generated by both V2V and V2I technologies. Other articles in the literature evaluated the exposure levels generated by V2V antennas at 5.9 GHz [16][17][18][21] and for the recent 5G-V2V technology operating at 3.5 GHz [19][20]. Specifically, the authors in [17][18] investigated the exposure levels generated by 5.9 GHz V2V technology through a deterministic approach on an adult [17] and children [18] placed in close proximity to the vehicle. The vehicle was equipped with two V2V antennas operating at 5.9 GHz with an input power of 30 dBm. By scaling the wbSAR values obtained in [17] with the maximum input power here used (33 dBm), the values obtained in [17] were almost the same as those obtained in the adults in our study. For the higher number of V2X antennas used in this study, we would expect higher wbSAR values than those obtained in [17]. However, there could be compensating effects between the number of antennas and the position of the human model, as in [17] the adult model was placed much closer to the vehicle than in our study. In [18], the authors compared the children exposure levels obtained with those of the adult obtained in [17]. As a result, they obtained the same evidence here found, i.e., for antennas mounted on the roof of the vehicle, the height of the human models was the parameter that mostly affected RF exposure levels on road users, where the adult models always received the greatest exposure compared to the children.

If we compared our results also with the dose absorbed by a passenger inside a vehicle, as calculated in [16], in our case we obtained wbSAR values that were slightly lower than those obtained in [16]. This is due to the different location of the antenna in [16]; indeed, compared to our case, placing the antennas on the side mirrors as in [16] results in a lower distance to the passenger head.

Differently from [16][17][18], the authors in [21], investigated the exposure levels generated by V2V technologies using an analytical approach to account for the variability of V2V exposure in urban scenarios of different characteristics. To provide a meaningful numerical comparison, we focused only on the data of [21] obtained from scenarios that look the most similar to our owns. In the scenarios of [21], among the human models investigated (both adults and children) the adults had the highest exposure levels with a maximum 99[th] percentile value that was much greater than that obtained here on the adults. As the analytical approach implemented by the authors [21] for computing the reflection and diffraction phenomenon was based on the same algorithms implemented on Wireless Insite [24] here used, the differences between the two studies were given mainly by the different characterization of the urban layout. Specifically in this study we modelled and simulated a realistic urban scenario (with roads, vehicles, buildings, trees, and grass), while in [21] the scenario was only characterized by factors that mimic an urban layout structure with only buildings and roads. Thus, on the contrary of [21], the data here obtained were evaluated considering the influence of the geometrical features of a realistic urban scenario.

Overall, compared with past studies, we discovered that the exposure levels generated by ITS-5.9 GHz technology in the outdoor urban scenario were generally lower than those obtained at a very close distance to the vehicle in free space [16][17][18] [21].

Currently, the IEEE 802.11p protocol [7] is the most used V2X communication protocol, but future research will focus on the innovative C-V2X protocol based on 5G technologies. For this reason, it would be interesting to provide a comparison between the exposure levels generated by the technologies here used, i.e., ITS-5.9 GHz based on IEEE 802.11p, with those obtained with the C-V2X protocol. From the literature, only one article [19] assessed the dose absorbed generated by the innovative V2V antennas at 3.5 GHz. In [19] the authors investigated the dose absorbed in the free space on an adult model positioned very close to the vehicle equipped with two 4x2 elements array antennas at 3.5 GHz. The wbSAR values found in this study [19] were lower than those found here. This result was quite expected as the patch array antenna performed beamforming capability and as such exposed less the human model to radiation compared to the V2V and V2I-5.9 GHz antennas which instead were omnidirectional antennas and spread radiation all over the azimuthal plane. As a result, 5G-V2V exposure levels, as well as those obtained in the current study, remain within the safety limits imposed by the ICNIRP [22] and IEEE [23] guidelines.

## 5. Conclusion

This article investigated for the first time the RF-EMF dose absorbed by road users in V2V and V2I realistic urban exposure. We found that adults always had exposure levels higher than children. The maximum wbSAR was $4.9 \cdot 10^{-4}$ W/kg which was well below the limits imposed by the basic restrictions of the ICNIRP and IEEE guidelines.

The exposure levels highly depended on the position of the road user, the size of the road user, and the 'objects' in the environment (i.e., the presence of buildings, vehicles, and vegetation like trees). The median wbSAR ranged from $10^{-7}$ W/kg to $10^{-5}$ W/kg across all the different exposure scenarios and human models investigated. The impact of V2I technologies on road users was revealed to be negligible compared to that of V2V technologies. We also found that the wbSAR values obtained in vehicular outdoor urban scenarios were generally lower than those obtained from other investigations inside the vehicle or outside the vehicle but at a closer position to the transmitting vehicles.

Funding

This work was financed by Project "EXPOAUTO - Cumulative real smart car exposure to radiofrequency electromagnetic fields in people of different ages from new technologies in automotive services and connected objects" [PNREST Anses, 2020/2 RF/05].

Acknowledge

The authors wish to thank Dr. Tarun Chawla from Remcom and his support/staff for providing support during the set-up of the simulations.